\documentclass[prl,twocolumn,amsmath,amssymb,aps,preprintnumbers,superscriptaddress]{revtex4-1}
\usepackage{physics}
\usepackage[normalem]{ulem}
\usepackage{bbold}
\usepackage{braket}
\usepackage{amsmath, amsfonts, amssymb}
\usepackage{graphicx}
\usepackage{float}
\usepackage{hyperref}
\usepackage[caption=false]{subfig}
\usepackage{mathtools}
\usepackage{tikz}
\newcommand*\dimer{\includegraphics[scale=0.18]{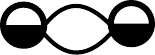}}
\newcommand*\bodw{\includegraphics[scale=0.5]{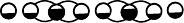}}
\usepackage{notes2bib}
\bibnotesetup{
	note-name = ,
	use-sort-key = false
}

\begin{document}
	\newcommand{\titleinfo}{Quantum Phases from Competing Van der Waals and Dipole-Dipole Interactions of Rydberg Atoms}
	\title{\titleinfo}
	
	\author{Zeki Zeybek}
	\email{zzeybek@physnet.uni-hamburg.de}
	\affiliation{The Hamburg Centre for Ultrafast Imaging, Universit{\"a}t Hamburg\\Luruper Chaussee 149, 22761 Hamburg, Germany}
	\affiliation{Zentrum f{\"u}r Optische Quantentechnologien, Universit{\"a}t Hamburg\\Luruper Chaussee 149, 22761 Hamburg, Germany}
	
	\author{Rick Mukherjee}
	\email{rmukherj@physnet.uni-hamburg.de}
	\affiliation{Zentrum f{\"u}r Optische Quantentechnologien, Universit{\"a}t Hamburg\\Luruper Chaussee 149, 22761 Hamburg, Germany}
	
	\author{Peter Schmelcher}
	\affiliation{The Hamburg Centre for Ultrafast Imaging, Universit{\"a}t Hamburg\\Luruper Chaussee 149, 22761 Hamburg, Germany}
	\affiliation{Zentrum f{\"u}r Optische Quantentechnologien, Universit{\"a}t Hamburg\\Luruper Chaussee 149, 22761 Hamburg, Germany}

	\begin{abstract}
		Competing short- and long-range interactions represent distinguished ingredients for the formation of complex quantum many-body phases. Their study is hard to realize with conventional quantum simulators. In this regard, Rydberg atoms provide an exception as their excited manifold of states have both density-density and exchange interactions whose strength and range can vary considerably. Focusing on one-dimensional systems, we leverage the van der Waals and dipole-dipole interactions of the Rydberg atoms to obtain the zero-temperature phase diagram for a uniform chain and a dimer model. For the uniform chain, we can influence the boundaries between ordered phases and a Luttinger liquid phase. For the dimerized case, a new type of bond-order-density-wave phase is identified.This demonstrates the versatility of the Rydberg platform in studying physics involving short- and long-ranged interactions simultaneously.
	\end{abstract}
	\maketitle
	
	\textit{Introduction.\textemdash} The interplay of short- and long-range interactions gives rise to diverse phenomena with implications in different areas such as study of electronic dynamics and stability in proteins \cite{Sheu,Gnandt,miyazawa_long-_2003,alshareedah_interplay_2019}, self-assembly in polymers \cite{patsahan_self-assembly_2021} and exotic quantum phases in condensed matter physics \cite{bacani_interplay_2017, igloi_quantum_2018, nishino_multistep_2019, azouz_competing_2022,zhu_quantum_2022}. However, the study of these phenomena in the natural biochemical and solid-state setups are challenging due to the limited control and the finite temperature environments. This has lead to a rapid growth in the use of ultra-cold systems for quantum simulation of many-body problems \cite{lewenstein_ultracold_2007, bloch_many-body_2008,bloch_quantum_2012}. These include the highly tunable short-range interactions with atoms in optical lattices \cite{landig_quantum_2016,gross_quantum_2017} to long-range interacting dipolar gases \cite{trefzger_ultracold_2011,baier_extended_2016}, polar molecules \cite{yan_observation_2013,hazzard_many-body_2014,Andris,WeimerPolar} and trapped ions \cite{monroe_programmable_2021, blatt_quantum_2012,Roy}. Realizing interactions with different scaling is of great interest not only to the broader condensed matter community, but also from a different viewpoint, it opens avenues for simulating chemical and biological processes that involve short- and long-range physics \cite{Sheu,Gnandt,miyazawa_long-_2003,alshareedah_interplay_2019,patsahan_self-assembly_2021}. This is particularly challenging since it requires implementing different power law interactions simultaneously, where other quantum simulating platforms such as trapped ions have limited utility.

	 %Although trapped ions have been used to simulate effective interactions that have power-law decay $1/r^{\alpha}$, such that $\alpha$ can be varied from $0$ to $ 3$, they can be remarkably sensitive to external fields and noise.

	\begin{figure}[t!]
		\includegraphics[width=\columnwidth]{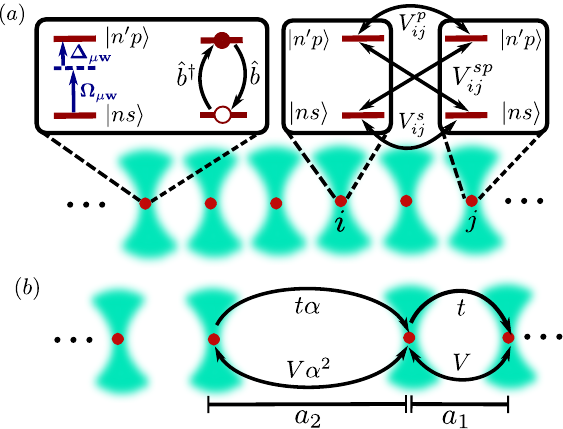}
		\caption{(a) Diagram depicting a uniform lattice of neutral atoms treated as two-level systems consisting of highly excited Rydberg states. Microwave laser with Rabi frequency $\Omega_{\mu w}$ and  detuning $\Delta_{\mu w}$ couples the levels. Atoms in the same Rydberg state experience vdW interactions with strengths $V^{s}_{ij}$, $V^{p}_{ij}$ while $V^{sp}_{ij}$ tunes the dipolar exchange interaction between different levels. The two-level system encodes the presence (absence) of a boson at a given site defined by $\hat b^\dagger (\hat b)$. (b) Dimerized chain with alternating intra-cell $a_1$ and inter-cell $a_2$ lattice constants with corresponding hopping ($t,t\alpha$) and off-site interactions ($V,V\alpha^2$).} 
		\label{fig:Setup}
	\end{figure}
	
	Platforms based on neutral Rydberg atoms have proven to be highly practical quantum simulators \cite{weimer_rydberg_2010,Mukherjee_2011,low_experimental_2012, browaeys_experimental_2016, morgado_quantum_2021} as their large dipole moments provide tunable strong interactions that range from dipole-dipole ($1/r^3$ scaling) to van der Waals ($1/r^6$ scaling). Rydberg-dressing \cite{Johnson2010, Rydberg-dressed} allows for a certain flexibility in controlling short- and long-range interactions simultaneously with applications in many-body physics  \cite{Henkel2012,Charge}, but they can be experimentally challenging to realize \cite{Balewski_2014,Zeiher}. Currently, the most common approach adopted in the context of quantum simulation with Rydberg atoms is to focus either on short-range (vdW) \cite{bernien_probing_2017, vdw_sylvain,ebadi_quantum_2021,verr,semeghini_probing_2021,scholl_quantum_2021,Sach2D,SachKag} or long-range (dipolar) interactions \cite{de_leseleuc_observation_2019,SPT_glass,dd_Les,dd_eisfeld}, but rarely both \cite{Forbid}. In the past, theoretical studies \cite{weimer_two-stage_2010,sela_dislocation-mediated_2011} were motivated by the limitations of the experiments to focus on either regime but never both. Only recently, experiments were performed with a pair of Rydberg states which potentially allow both ranges of interactions \cite{de_leseleuc_observation_2019,scholl_microwave_2022} but were never comprehensively exploited.

	%In the past, theoretical studies [Phys. Rev. Lett. \textbf{105},230403 (2010), Phys. Rev. A \textbf{104}, 063301 (2021)] were motivated by the limitations of the experiments to focus on either regime but never both. Only recently, experiments were performed with a pair of Rydberg states which potentially allow both ranges of interactions [Science \textbf{365}, 775 (2019), PRX Quantum \textbf{3}, 020303 (2022)] but were never sufficiently exploited.
	
	%However, most of the applications of quantum simulation with Rydberg atoms have focused on either exploiting the van der Waals \cite{bernien_probing_2017, vdw_sylvain,ebadi_quantum_2021,semeghini_probing_2021,scholl_quantum_2021,Sach2D,SachKag} or the dipole-dipole interaction \cite{de_leseleuc_observation_2019,SPT_glass,dd_Les,dd_eisfeld}, but rarely both together.

	In this letter, we propose an alternative approach to study short-long range physics by combining the effects of van der Waals and dipole-dipole interactions of Rydberg atoms. Using the one-dimensional (1D) uniform/dimerized lattices, we study the ground state phase diagram and unveil the flexibility in accessing different regimes of the ordered and liquid phases. For the uniform chain, the competition between the interactions is reflected in the competing boundaries between the gapless Luttinger liquid (LL) and the gapped density-wave (DW) ordered phases. In the dimerized chain, apart from realizing the individual phases of bond-order (BO) and DW, we find unique bond-order-density-wave (BODW) phase that has not been previously explored using conventional dimerized model \cite{BOW,hayashi_competing_2022}. 
	
	\textit{Model and Hamiltonian}.\textemdash We discuss the Rydberg setup and its mapping to extended Bose-Hubbard model which distinguishes itself from existing bosonic models \cite{Rossini_2012,Maik_2013,Sach2D,SachKag}. The uniform and dimerized lattices are considered where the latter is known for rich physics involving topological and insulating phases \cite{su_solitons_1979, chen_quantum_2010,Topo_dim,Phase_dim,fraxanet_topological_2022, hayashi_competing_2022}.
	
	As illustrated in Fig.~\ref{fig:Setup}, the setup consists of a linear chain of trapped atoms which could either have uniform or dimerized lattice configurations. Each atom is a two-level system made of $\ket{ns}$ and $\ket{n^{\prime}p}$ Rydberg states, where $n,n'$ are principal quantum numbers. Unlike most Rydberg simulators with one ground state and a Rydberg level, the pair of Rydberg states considered here allows the system to have two types of interactions which differ in range and character: (i) the short-range van der Waals (vdW) interactions between the $ns-ns$ and $n^{\prime}p-n^{\prime}p$ and (ii) the long-range dipolar interaction which causes a state exchange between atoms in different Rydberg levels $ns-n^{\prime}p$. The Rydberg interactions along with the microwave laser coupling between $\ket{ns}$ and $\ket{n^{\prime}p}$ levels are all schematically shown in Fig.~\ref{fig:Setup}(a). The corresponding atomic Hamiltonian describing the full setup with uniform lattice spacing $a$ is given as
	\begin{align}\label{eqn:Atomic_Ryd}
		\hat H_A &= \sum_i \Big[\frac{\Omega_{\mu w}}{2}(\hat\sigma^{sp}_{i} + \hat\sigma^{ps}_{i}) - \Delta_{\mu w}\hat\sigma^{pp}_{i} \Big] + V^{p}\sum_{i<j} \frac{\hat\sigma^{pp}_{i}\hat\sigma^{pp}_{j}}{\abs{i-j}^6} \notag \\
		&+ V^{s}\sum_{i<j} \frac{\hat\sigma^{ss}_{i}\hat\sigma^{ss}_{j}}{\abs{i-j}^6}  
		+V^{sp}\sum_{i<j} \Big(\frac{\hat\sigma^{sp}_{i}\hat\sigma^{ps}_{j}}{\abs{i-j}^3}+ \text{h.c.}\Big) .
	\end{align}
	Here $\hat\sigma^{\alpha \beta}_i = \ket{\alpha}_i\bra{\beta}$ is the projection operator to the relevant atomic state with $\alpha,\beta \in \{\ket{ns},\ket{n^{\prime}p}\}$ at site $i$. $V^{p}=C^{p}_6/a^6$ and $V^{s}=C^{s}_6/a^6$ are the strength of vdW interactions, where $C^{s}_6$ and $C^{p}_6$ are the dispersion coefficients. $V^{sp}=C_3/a^3$ is the dipole-dipole interaction strength with $C_3$ as the exchange coefficient. There are experimental realizations of the above Hamiltonian \cite{de_leseleuc_observation_2019, scholl_microwave_2022}. 
	
	In order to represent Eq.~\ref{eqn:Atomic_Ryd} in the Bose-Hubbard picture, the occupation of state $\ket{n^{\prime}p}$ at site $i$ is associated with the presence of a boson at that site and denoted by $\ket{\bullet}_i$ while $\ket{\circ}_i$ means the absence of a boson which implies the  occupation of state  $\ket{ns}$.  With these definitions, an arbitrary state  $\ket{ns~n'p~n'p~ns \dots}$ is written as $\ket{\circ \bullet \bullet \circ \dots}$. Since each atom cannot have more than one excitation $\ket{n^{\prime}p}$, having two particles at the same site is prohibited, which imposes a hard-core constraint. Defining the $\hat b^\dagger (\hat b)$ as the bosonic creation (annihilation) operator, $\hat H_A$ is re-written as follows,
	\begin{align}\label{eqn:EBHM_Ryd} 
		\hat H_{eBH} &= \sum_{i<j} t_{ij}( \hat b^{\dagger}_i \hat b_j +\text{h.c.}) + \sum_{i<j}V_{ij} \hat n_i \hat n_j \notag \\ 
		& -  \sum_{i}(\Delta_{\mu w} + \mathcal{I}_i ) \hat n_i +\frac{\Omega_{\mu w}}{2}\sum_{i}(\hat b^{\dagger}_i+ \hat b_i) ,
	\end{align}
	where we used the mapping $\hat\sigma^{ps} \rightarrow \hat b^{\dagger}$, $\hat\sigma^{pp} \rightarrow \hat n=\hat b^{\dagger}\hat  b$ and $\hat\sigma^{ss} \rightarrow \mathbb{1}- \hat n$ with $(\hat b_i^{\dagger})^2=0$. The first term in Eq.~\ref{eqn:EBHM_Ryd} is the long-range hopping $t_{ij} = V^{sp}/\abs{i-j}^3$ which is encoded by the dipolar exchange interaction. $V_{ij} = V/\abs{i-j}^6$ is the repulsive off-site density interaction, where $V = V^{s}+V^{p} = C_6/a^6$ and $C_6=C^s_6+C^p_6$ is the combined dispersion coefficient. The chemical potential $(\Delta_{\mu w} + \mathcal{I}_i)$ determines the density of excitations $\ket{n^\prime p}$ (number of bosons) in a lattice. The site-dependent contribution $\mathcal{I}_i=\sum_{i\neq j}\frac{V^{s}}{\abs{i-j}^6}$ is an energy offset for a fixed value of the chemical potential and can be ignored in the bulk, thus $\mu_i \rightarrow \mu = \Delta_{\mu w}$. The $\hat H_{eBH}$ differs from other extended Bose-Hubbard models \cite{Rossini_2012,Maik_2013,Sach2D,SachKag} in several aspects: $(i)$ the existence of longer-range hopping and interactions and $(ii)$ the last term in $\hat H_{eBH}$ breaks the global U$(1)$ symmetry causing the number of bosons to be a non-conserved quantity. These aspects will play a role in the phase diagrams obtained later.
	\begin{figure*}[t!]
		\includegraphics[width=\textwidth]{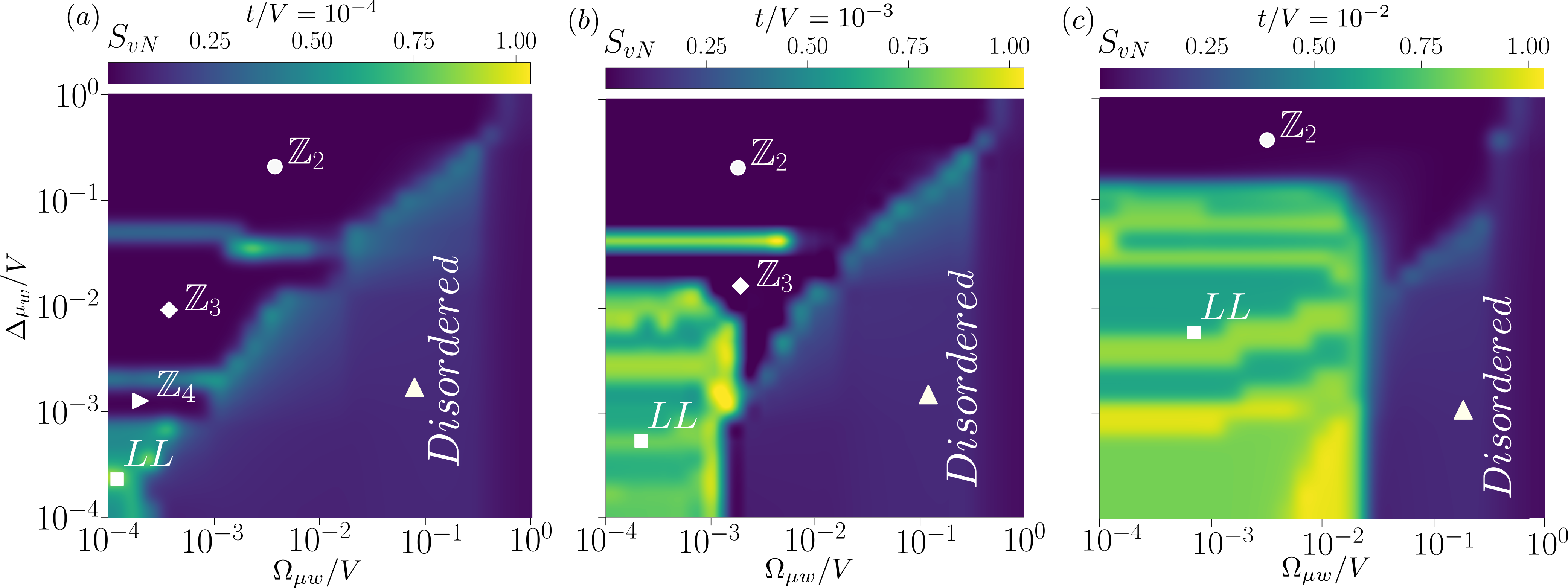}
		\caption{Phase diagrams showing the ground-state entanglement entropy $S_{vN}$ of $\hat H_{eBH}$ in the $(\Delta_{\mu w},\Omega_{\mu w})$ parameter space for system size $L=121$ with varying $t/V$ in (a), (b) and (c) respectively. The dark-shaded blue lobes in the top left part of the phase diagrams represent a vanishing $S_{vN}$ which correspond to different gapped ordered phases $Z_{q=2,3,4}$. The yellow-green regions represent finite $S_{vN}$ corresponding to the gapless Luttinger liquid (LL) phase. For large values of $\Omega_{\mu w}$, one obtains the disordered phase which is shown as light-shaded blue. Verification of the individual phases is provided in \cite{SM}.}
		\label{unif_PD}
	\end{figure*}
	
	Fig.~\ref{fig:Setup}(b) depicts the dimerized configuration formed by two sub-lattices with alternating lattice constants $a_1$ and $a_2$. The dimerized version of Eq.~\ref{eqn:EBHM_Ryd} is provided in \bibnote[SM]{See Supplemental Material for (1) Physical parameters for the realization of the setup, (2) Mapping of the atomic Hamiltonian to extended Bose-Hubbard model including the dimerized case, (3-4) Additional analysis on verification of individual phases and (5) Numerical method details, which includes Refs. \cite{saffman_quantum_2010,leseleuc_de_kerouara_quantum_2018, sibalic_arc_2017,weber_calculation_2017,BBR1,BBR2,PRA1,hauschild_efficient_2018}}, whose many-body energy spectrum for $\Omega_{\mu w}=0$ constitutes of many distinct manifolds, each of which is characterized by a fixed number of bosons. For large negative values of the microwave detuning $\Delta_{\mu w}$, one obtains a completely empty lattice (all atoms in $\ket{ns}$ state). As $\Delta_{\mu w}$ increases, the number of bosons added to the lattice also increases. Similar to experiment \cite{de_leseleuc_observation_2019}, an adiabatic sweep through the parameters $(\Omega_{\mu w}(t),\Delta_{\mu w}(t))$ can take the lattice system of size $L$ from one manifold with zero bosons to another manifold of $N$ bosons giving a filling $\rho=N/L$. After reaching a given filling $\rho$, the microwave laser is switched off and the following Hamiltonian is written as
	\begin{align}\label{eqn:EBHM_Ryd_Dim}
		\hat H_{dim} &= t\sum_{i\in odd} ( \hat b^\dagger_i \hat b_{i+1} + \text{h.c.}) + t\alpha\sum_{i\in even} (\hat b^\dagger_i \hat b_{i+1} + \text{h.c.}) \notag \\&+ V\sum_{i\in odd} \hat n_i \hat n_{i+1} + V\alpha^2\sum_{i\in even} \hat n_i \hat n_{i+1} + \hat H_{LR} .
	\end{align}
	The even and odd sums represent the intra- and inter-cell terms respectively and the dimerization constant $\alpha=(a_1/a_2)^3$ controls the degree of dimerization in the lattice. $t=-C_3/a_1^3$ and $V=C_6/a_1^6$ are the intra-cell hopping and off-site interaction strengths respectively with $C_6,C_3>0$. $\hat H_{dim}$ deviates from existing dimer models \cite{Topo_dim,Phase_dim,fraxanet_topological_2022,hayashi_competing_2022} in the sense that it has both local as well as long-range hopping and  off-site interaction defined under $\hat H_{LR}$ (explicitly given in \cite{SM}). The fact that it has dimerization in interaction and not just in hopping will play a crucial role in the phase diagrams.
	
\begin{figure*}[t!]
		\includegraphics[width=\textwidth]{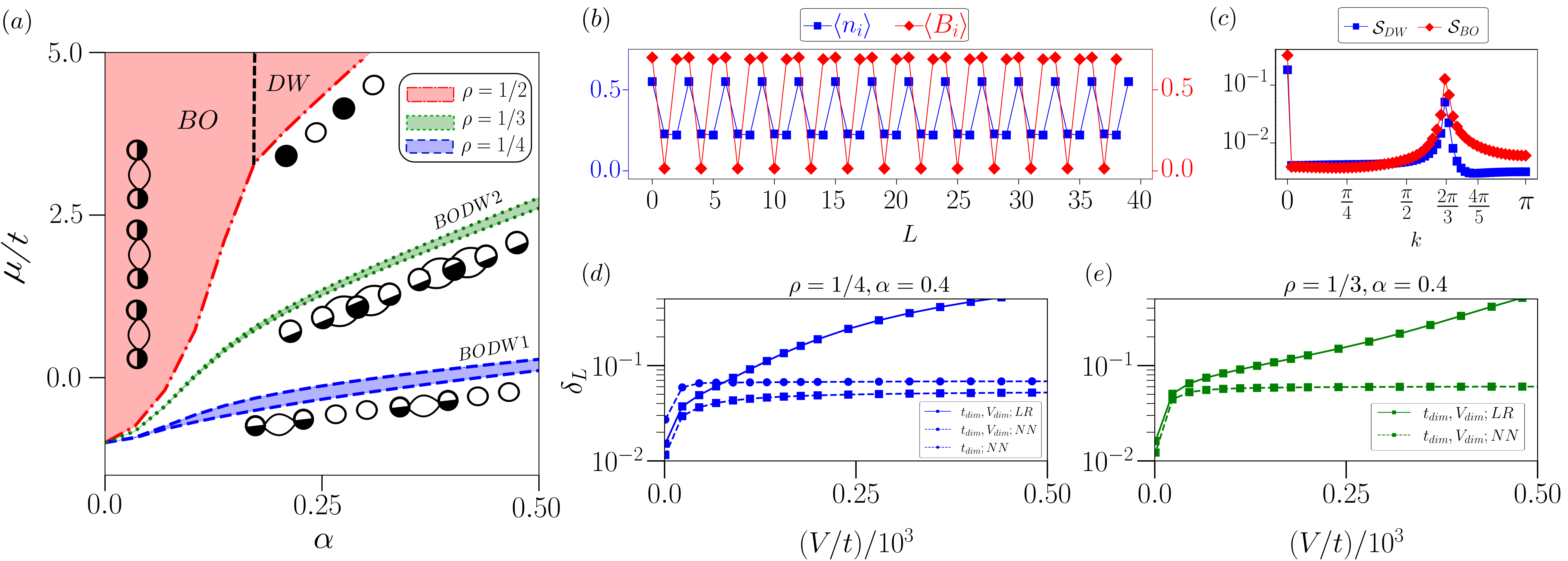}
		\caption{(a) Gapped phases of $\hat H_{dim}$ are shown as a function of $\alpha$ with fixed $V/t=200$ for system size $L=240$. The red dashed-dotted line defines the lower boundary at $\rho=1/2$ with the vertical dashed black line that separates the BO and DW regions. The green dotted and the blue dashed lines determine the boundaries of the BODW phases at $\rho=1/3$ and $\rho=1/4$ respectively. Density and bond formation in each phase are symbolically represented with partially filled circles (superposition of $ns$ and $n'p$ states)  and curved lines between sites respectively. (b) Expectation values of the bond energy (red, diamond) and density (blue, square) operators for the BODW phase at $\rho=1/3$ for $\alpha=0.4$ and $V/t=200$ is displayed. The corresponding structure factors are shown in (c). (d,e) Figure comparing the gap $\delta_L$ for BODW phases at filling $\rho=1/4$ and $\rho=1/3$ is shown with different types of couplings in Eq.~\ref{eqn:EBHM_Ryd_Dim}. The gap $\delta_L$ is plotted as a function of $V/t$ with fixed $\alpha$ and system size $L=240$. Cases with (squares) and without (circles) dimerization in the interaction are considered for different range of interactions: nearest-neighbour (dashed line with NN) and long-range (solid line with LR).}
		\label{dim_PD}
	\end{figure*}

	\textit{Results.\textemdash} Figures~\ref{unif_PD} and  \ref{dim_PD} are the ground-state phase diagrams for $\hat H_{eBH}$ and $\hat H_{dim}$ obtained using finite-size DMRG \cite{white_density_1992,white_density-matrix_1993}. More details about numerics are in \cite{SM}. In earlier works for a uniform lattice \cite{weimer_two-stage_2010, sela_dislocation-mediated_2011, bernien_probing_2017, yu_fidelity_2022}, one finds the LL phase to be always dominated by the ordered phases for the entire region of allowed laser parameters. In contrast, here we show that the boundaries between ordered and LL phases are easily adjustable, and find scenarios where LL even dominates. This is possible due to the competition of vdW and dipolar exchange terms. The same flexibility in the boundaries of the BODW phases are seen in the dimerized case. Moreover, the existence of BODW phase for $\rho = 1/3$ filling is shown, which does not occur in conventional models \cite{BOW,hayashi_competing_2022}. 
	
	In Fig.~\ref{unif_PD}(a-c), we compute the half-chain bipartite entanglement entropy $S_{vN}\equiv -\Tr(\rho_r\ln{\rho_r})$ of the ground state over the parameter space $(\Omega_{\mu w},\Delta_{\mu w})$ for fixed hopping $t$, where $\rho_r$ is the reduced density matrix of half of the chain. This has been performed for a varying relative strength of the hopping $t$ as shown in Fig.~\ref{unif_PD}(a-c). DW are many-body ground states that are ordered (crystalline) and are characterized with unit cell $p/q$ where $p$ denotes the number of bosons and $q$ is the size of the unit cell. For example, the circle markers in Fig.~\ref{unif_PD}(a-c) correspond to phases that break $\mathbb{Z}_2$ translational symmetry with $p/q=1/2$. $\mathbb{Z}_2$ phase is described by the state $\ket{\bullet \circ \bullet \dots \bullet \circ \bullet}$, which is a product state and thus possesses a vanishing $S_{vN}$. Similarly, higher period DW phases ($\mathbb{Z}_{q=3,4}$) are also shown in Fig.~\ref{unif_PD}(a) and (b).
	
	Although both the hopping term $t$ and Rabi coupling $\Omega_{\mu w}$ introduce quantum fluctuations to the system, they have different effects on the ordered states. For example, if $\Omega_{\mu w} \gg t$, then one obtains a disordered state. However when $\Omega_{\mu w}$ becomes comparable to $t$, then we have either an ordered phase or a LL phase depending on the value of $\Delta_{\mu w}$. Close to the classical regime ($\Omega_{\mu w}=t\simeq0$), the range of the ordered phase $\mathbb{Z}_q$ in terms of the detuning is given as $\delta \Delta_{\mu w} \sim V_{i,i+q-1} + \mathcal{O}(V_{i,i+q})$ \cite{bak_one-dimensional_1982, sela_dislocation-mediated_2011}. Thus for low values of $(t,\Omega_{\mu w})$, one finds a host of ordered phases $\mathbb{Z}_{q=2,3,4}$ as seen in Fig.~\ref{unif_PD}(a). As $t$ increases such that $t\ge V_{i,i+q-1}$, then ordered phases with unit cells larger than $q$ get washed out and instead the LL phase takes over as seen in Fig.~\ref{unif_PD}(b-c). This condition is satisfied as the vdW interaction has the combined effect of $ns$ and $n' p$ states for different $n$ and $n'$  \cite{SM}. Universal properties of the LL phase such as power-law decay of correlations and the central charge $c=1$ \cite{calabrese_entanglement_2009,pollmann_theory_2009} have been verified \cite{SM}.
	
	Figure \ref{dim_PD}(a) is obtained by determining the single-particle excitation gap $\delta_{L}(\alpha,V)= \mu^{+}(\alpha,V) - \mu^{-}(\alpha,V)$ as a function of $\alpha$ with fixed $V$. Thus, the extent of the gapped phases in the phase diagram scales as $\delta_{L}$. Here $\mu^+ = E(N+1) - E(N)$ and $\mu^- = E(N) - E(N-1)$ are the chemical potentials that define the boundaries of the gapped phases for a given  filling $\rho$ and $E(N)$ is the ground state energy for a system of $N$ bosons defined by $\hat H_{dim}$. In Fig.~\ref{dim_PD}(a), four types of gapped phases (DW, BO, BODW1, BODW2) are obtained for different values of filling $\rho$ in the $(\alpha, \mu)$ parameter space with constant $V$. DW are the ordered phases as discussed before while the BO phase is a product of independent dimers which is represented as $\prod\limits_{i=0} (\frac{\hat b^{\dagger}_{2i}+\hat b^{\dagger}_{2i+1}}{\sqrt{2}})\ket{\circ \circ \dots \circ}=\ket{\dimer \hspace{0.8mm}\dimer \dots}$, where each dimer corresponds to two sites sharing a single delocalized boson (\dimer). Bond-order-density-wave (BODW) has the characteristic of both bond ordering and density wave ordering. Numerical verification of the individual phases is provided in \cite{SM}.
	
	In Fig.~\ref{dim_PD}(a), at $\rho=1/2$, the gap remains open for all values of $\alpha$ and hosts two different ordered phases, BO and DW. Low values of $\alpha$ is indicative of a highly dimerized lattice where the nearest-neighbour processes within a unit cell dominates over long-range processes such as inter-cell hopping and extended off-site interactions. At $\rho=1/2$ filling, this means significant energy costs in adding/removing bosons which leads to a region of energy-gap corresponding to the BO phase as seen in Fig.~\ref{dim_PD}(a). As $\alpha$ is increased, the long-range effects of hopping and interaction become relevant. But if $V/t$ is sufficiently large, which is the case in Fig.~\ref{dim_PD}(a), then the repulsive vdW interaction leads to a DW phase for the $\rho=1/2$ filling.
	
	For any other filling $\rho \neq 1/2$, the gap closes as $\alpha \rightarrow 0$ as seen in Fig.~\ref{dim_PD}(a), which implies that there is no energy cost in adding/removing bosons and allowing free movement of bosons across the lattice (LL phase) is favored over the BO phase as seen for $\rho = 1/4, 1/3$ fillings. As $\alpha$ increases, long-range processes become dominant and for sufficiently large $V/t$, BODW phases are obtained for $\rho=1/4, 1/3$ fillings in contrast to the DW phase that we get for $\rho=1/2$ filling. BODW phases arise from the cumulative effect of dominant long-range repulsive interactions at large values of $\alpha$ and the constraint of sharing certain number of bosons among the lattice due to the fixed filling fraction. However, it is sufficient to have nearest-neighbor interactions to stabilize the BODW1 phase \cite{hayashi_competing_2022}, which consists of dimers in every alternating unit cells. Thus, the BODW1 phase can be described by a product state $\prod\limits_{i=0} (\frac{\hat b^{\dagger}_{4i}+ \hat b^{\dagger}_{4i+1}}{\sqrt{2}})\ket{\circ \circ \dots \circ}=\ket{\dimer \circ \circ \hspace{0.8mm} \dimer\circ \circ \dots}$ to a good approximation. Unlike BODW1, long-range interactions are needed to stabilize the BODW2 phase. Therefore, it cannot be described with independent dimers but rather is depicted as $\ket{\bodw \dots}$, where a pair of dimers are shared between three sites. The latter BODW2 phase has not been explored before.
	
	%On the other hand, the BODW2 phase for $\rho=1/3$ is depicted as $\ket{\bodw \dots}$, where a pair of dimers are shared between three sites. \textcolor{blue}{Therefore, it cannot be described by tiling the lattice with independent dimers as it is the case for BODW1 and BO phases.} 
	
	In Figs.~\ref{dim_PD}(b,c), the characterization of the BODW phase at $\rho=1/3$ is shown \cite{SM}. The BO nature is probed with the bond order structure factor $\mathcal{S}_{BO}(k) = (1/L^2) \sum_{i,j}e^{ikr}\braket{\hat B_i \hat B_j}$, where $\hat B_i = \hat b^\dagger_i \hat b_{i+1} + \text{h.c.}$ is the bond energy operator, while the density wave structure factor is $\mathcal{S}_{DW}(k) = (1/L^2) \sum_{i,j}e^{ikr}\braket{\hat n_i \hat n_j}$. In Fig.~\ref{dim_PD}(b), oscillations in $\hat n_i$ implies that the bosons primarily occupy every third site on the chain (analogous to $\mathbb{Z}_{q=3}$), thus giving the DW character of the phase. BO oscillations point to a state with $p=2$ bonds for a unit cell of size $q=3$ as two bonds form among three sites. These findings are also reflected in the peaks of the structure factors $\mathcal{S}_{DW}(k)$ and $\mathcal{S}_{BO}(k)$  at $k=2\pi/3$ as shown in Fig.~\ref{dim_PD}(c). Figure \ref{dim_PD}(d) shows the gap $\delta_L$ for the BODW1 phase as a function $V/t$ with fixed dimerization $\alpha=0.4$ for different cases. One finds large energy gaps for the model with dimerized long-range interactions when compared to the almost vanishing gap for the non-dimerized and nearest-neighbour dimerized interacting models. A similar analysis applies to BODW2 in Fig.~\ref{dim_PD}(e). The key feature that is required to obtain the BODW2 phase is the necessity of beyond nearest-neighbour contributions along with dimerization in the interaction. However, the stability of the phases and thus their boundaries depend on the scaling of the interactions with distance. This highlights the key merits of our setup when compared to existing dimer models where only the hopping term is dimerized \cite{chen_quantum_2010,Topo_dim,Phase_dim,fraxanet_topological_2022, hayashi_competing_2022}. 
	
	For the Rydberg states considered in this work, we have a lifetime of few hundred of microseconds. This implies that for a chain of $10-20$ atoms, the system lifetime is on the order of few tens of microseconds which is sufficient for the phases to be experimentally realized while taking into account the relevant dissipative processes. A detailed analysis on the experimental feasibility is provided in \cite{SM}.
	
	%Experimentally relevant parameters to observe these phases are discussed in \cite{SM}. 
	
	%In particular, we provide a detailed analysis on the timescales required to observe these phases experimentally.

	\textit{Conclusion and outlook.\textemdash} Many-body systems with interactions operating over different length scales host a wide range of phenomena in nature. This work promotes the quantum simulation of such phenomena using Rydberg atoms where the interplay between vdW and dipolar interactions provide a long-range dimerized Hubbard model. The ground state phase diagram of this model in 1D is characteristically distinct from conventional models in two key aspects: larger tunability for the LL phase in the uniform case and the existence of a novel BODW phase in the dimerized case. Future works will include the investigation of higher dimensional lattices, different geometries and out-of-equilibrium dynamics with the recently developed multilayer multiconfigurational approach for spin systems \cite{Fabian}.

	\begin{acknowledgments}
		This work is funded by the Cluster of Excellence
		``CUI: Advanced Imaging of Matter'' of the Deutsche Forschungsgemeinschaft (DFG) - EXC 2056 - Project ID 390715994. This work is funded by the German Federal Ministry of Education and
Research within the funding program ``quantum technologies - from
basic research to market" under contract 13N16138.
	\end{acknowledgments}
	
	\bibliographystyle{apsrev4-1}
	\bibliography{main_ref}

\clearpage
\onecolumngrid
\newpage

\setcounter{equation}{0}            % reset equation counter
\setcounter{section}{0}    % reset section counter
\setcounter{figure}{0}    % reset section counter
\renewcommand\thesection{\arabic{section}}    % puts letters as section numbering
\renewcommand\thesubsection{\arabic{subsection}}    % puts letters as section numbering
\renewcommand{\thetable}{S\arabic{table}}
\renewcommand{\theequation}{S\arabic{equation}}
\renewcommand{\thefigure}{S\arabic{figure}}
\setcounter{secnumdepth}{2}
\setcounter{page}{1}

\begin{center}
	{\large Supplemental Material: \\ 
		\titleinfo}
\end{center}
\thispagestyle{empty}

\section{Experimental parameters for Rydberg models}
In this section, the experimental realization of the Rydberg Hamiltonian is discussed. In the main article, $\hat H_A$ (see Eq. 1 in the main article) is a collection of two-level systems consisting of Rydberg states $\{\ket{r},\ket{r^{\prime}}\}$. As a precursor, the setup will have all atoms in their electronic ground state $\ket{gg\dots g}$  which is coupled to the Rydberg level $\ket{rr\dots r}$ using either a single-photon or two-photon excitation scheme \cite{browaeys_experimental_2016,low_experimental_2012} with effective Rabi frequency $\Omega$ and effective detuning $\Delta$. The Hamiltonian describing the precursor setup is as follows,
\begin{equation}
	\hat H_0 = \frac{\Omega}{2}\sum (\ket{r}_i\bra{g}+\ket{g}_i\bra{r})-\Delta\sum_i \ket{r}_i\bra{r} + \sum_{i,j} V_{ij}(\ket{r}_i\bra{r} \otimes \ket{r}_j\bra{r}), 
\end{equation}
where $V_{ij} =C^{rr}_6/(a\abs{i-j})^6$ is the interaction between two atoms in Rydberg state $\ket{r}$ (which in our case is the $nS$ state) and $a$ is the lattice constant. Since the initial state of our system, for both uniform and dimerized cases involves having all atoms in the Rydberg nS state, the lattice spacing $a$ must be chosen larger than the blockade radius, $r_b=(C^{nS-nS}_6/2\Omega)^{1/6}$. Another aspect, with regards to experimental realization is the issue of initial state preparation. All atoms should be transferred to the $ns$ state which sets a minimum bound on the Rabi frequency $\Omega$ as it needs overcome the blockade regime. For typical two-photon excitation of the ground to the Rydberg state, this is not an issue. However, in the such two-photon excitation schemes, one needs to be careful about the lifetime of intermediate states  \cite{vdw_sylvain}. Other issues such as imperfect lattice filling and atom losses although challenging, are regularly tackled in current experiments \cite{bernien_probing_2017}.

But large values of lattice spacing will imply that dipole-dipole interactions will be dominant whereas in this work, we are mainly interested in achieving $V \gg t$ for finite values of $t$. For this purpose, we choose different principal quantum numbers for the two Rydberg states, $\ket{r}= \ket{nS}$ and $\ket{r'}= \ket{n^{\prime}P}$ and take advantage of the different scaling laws $V \propto n^{11}/a^6$ and $t \propto n^{4}/a^3$. The Hamiltonian $\hat H_A$ is realized by coupling two Rydberg levels $\ket{r}\leftrightarrow \ket{r^{\prime}}$ with a microwave laser with parameters $\Omega_{\mu w}$ and $\Delta_{\mu w}$ as mentioned in the main article. Assuming a Rabi frequency $\Omega=100$ MHz the two-levels $\{\ket{60S_{1/2},1/2},  \ket{61P_{1/2},-1/2}\}$ with $a\in [4,7]$ $\mu$m can lead to $t/V \in [10^{-2},10^{-1}]$. By increasing $n$, even lower $t/V$ can be obtained. The setup $\{\ket{90S_{1/2},1/2},  \ket{91P_{1/2},-1/2}\}$ with $a\in [9,19]$ $\mu$m can roughly provide $t/V \in [10^{-3},10^{-1}]$. For purposes of generality, in table \ref{tab:ex} we show other values of dispersion coefficients for Rb atoms which may also be useful for this work. Our model requires the vdW interactions to be repulsive which occurs for  $nS$ states \cite{saffman_quantum_2010} and $nP$ states with $n>42$ \cite{leseleuc_de_kerouara_quantum_2018}.
	\begin{table}[H]
		\caption{\label{tab:ex}Dispersion coefficients for different $n, n'$ for $\prescript{87}{}{\text{Rb}}$ using \cite{sibalic_arc_2017,weber_calculation_2017}.}
		\begin{ruledtabular}
			\begin{tabular}{lllll}
				$\ket{\downarrow}$ & $\ket{\uparrow}$ & $C_3[GHz.\mu m^3]$ & $C_6^{s}[GHz.\mu m^6]$ & $C_6^{p}[GHz.\mu m^6]$ \\
				$\ket{60S_{1/2},1/2}$ &$\ket{59P_{1/2},-1/2}$ &$2.51$ & $135.29$&$1.89$\\
				$\ket{60S_{1/2},1/2}$ &$\ket{60P_{1/2},-1/2}$ &$3.04$ & $135.29$&$2.68$\\
				$\ket{60S_{1/2},1/2}$ &$\ket{61P_{1/2},-1/2}$ &$0.04$ & $135.29$&$3.15$\\
				$\ket{90S_{1/2},1/2}$ &$\ket{89P_{1/2},-1/2}$ &$13.85$ & $16500.87$&$521$\\
				$\ket{90S_{1/2},1/2}$ &$\ket{90P_{1/2},-1/2}$ &$16.35$ & $16500.87$&$597$\\
				$\ket{90S_{1/2},1/2}$ &$\ket{91P_{1/2},-1/2}$ &$0.23$ & $16500.87$&$682$\\
			\end{tabular}
		\end{ruledtabular}
	\end{table}

The relevant dissipation and dephasing processes for our system are the following:
\begin{itemize}
\item Spontaneous decay of $ns$ and $n^{\prime}p$ states: For an archetypical two-level Rydberg system $\{\ket{90S_{1/2},1/2}$, $\ket{91P_{1/2},-1/2}\}$, we have a lifetime $\tau \sim 800$ $\mu$s. A many-body system size of $N=20$ atoms, the system lifetime is $\tau_{chain} \sim 40 $ $\mu$s which sets an upper bound to the time window for the experiments to observe the phases.	

\item Black-body radiation effects: For typical ambient temperatures $T= 70-300$ K, one obtains an effective Rydberg lifetime including black-body effects around $\sim 200-500$ $\mu$s ($80S$) and $\sim 300-500$ $\mu$s ($80P$) \cite{BBR1,BBR2,PRA1}. The effective lifetimes are reduced due to black-body effects but can be mitigated by having lower temperatures.

\item Shot-to-shot fluctuations in atomic positions:  This gives rise to interaction-induced dephasing, especially for very small inter-atomic spacings. In summary, for large enough lattice spacings (away from F{\"o}rster resonances), we can ignore these effects. Moreover, for the typical trap frequencies available in current experiments, the marginal fluctuations in the interactions do not seem to affect the robustness of the phase provided the laser parameters are chosen appropriately.
\end{itemize}
In order to estimate the required time scales to observe phases where hopping processes are relevant,  we compare the system lifetime $\tau_{chain}$ against the inverse of the hopping term. For, $\{\ket{90S_{1/2},1/2}$, $\ket{91P_{1/2},-1/2}\}$, we have $C_3/h = 0.23$ [GHz.$\mu m^3$] and thus the coupling ratio in the  range of $t/V \in [10^{-3},10^{-1}]$ for a lattice constant $a$ of few $\mu$m. In such a system, the analysis yields hopping time scales in the range of $[33, 3.3]$ $\mu$s depending on the desired $t/V$ which is well below $\tau_{chain} \sim 40 $ $\mu$s. Of course, to study these phases for larger systems can be an issue.

\section{Boson mapping of Rydberg Hamiltonian to uniform and dimerized models}
Here the hard-core boson mapping of $\hat H_A$ to $H_{eBH}$ is discussed. The atomic Hamiltonian $\hat H_A$ as given in the main article is
\begin{equation*}
	\label{eqn:H_A}
	\hat H_A = \frac{\Omega_{\mu w}}{2}\sum_i(\hat\sigma^{sp}_{i} + \hat\sigma^{ps}_{i}) - \Delta_{\mu w}\sum_i\hat\sigma^{pp}_{i} 
	+ V^{p}\sum_{i<j} \frac{\hat\sigma^{pp}_{i}\hat\sigma^{pp}_{j}}{\abs{i-j}^6} 
	+ V^{s}\sum_{i<j} \frac{\hat\sigma^{ss}_{i}\hat\sigma^{ss}_{j}}{\abs{i-j}^6}  
	+V^{sp}\sum_{i<j} \Bigg(\frac{\hat\sigma^{sp}_{i}\hat\sigma^{ps}_{j}}{\abs{i-j}^3}+ \text{h.c.}\Bigg).
\end{equation*}
Identifying boson creation with the excitation of atoms to $ \ket{n^{\prime}p} $ levels we can reformulate the problem in terms of the extended Bose-Hubbard model in which we deal with hard-core bosons. Making use of the transformation $\hat\sigma^{ps} \rightarrow \hat b^{\dagger}$, $\hat\sigma^{pp} \rightarrow \hat n=\hat b^{\dagger}\hat \hat b$, $\hat\sigma^{ss} \rightarrow \mathbb{1}- \hat n$,  $\mathbb{1}=\ket{ns}\bra{ns} + \ket{n^{\prime}p}\bra{n^{\prime}p}$ and collecting the common terms lead to,
\begin{align}
	\hat H_{eBH} &= \frac{\Omega_{\mu w}}{2}\sum_i (\hat b^{\dagger}_i+\hat b_i) - \Delta_{\mu w}\sum_i \hat n_i + V^p\sum_{i<j} \frac{\hat n_{i}\hat n_{j}}{\abs{i-j}^6} + V^s \sum_{i<j}  \frac{(1-\hat n_{i})(1-\hat n_{j})}{\abs{i-j}^6} \notag \\ \notag  &+V^{sp}\sum_{i<j} \Bigg(\frac{\hat b^{\dagger}_{i}\hat b_{j}}{\abs{i-j}^3}+ \text{h.c.}\Bigg) \\ \notag 
	&= \frac{\Omega_{\mu w}}{2}\sum_i (\hat b^{\dagger}_i+\hat b_i) - \Delta_{\mu w}\sum_i \hat n_i + (V^p+V^s)\sum_{i<j} \frac{\hat n_{i}\hat n_{j}}{\abs{i-j}^6} - \sum_i \Bigg(\sum_{i\neq j}\frac{V^{s}}{\abs{i-j}^6}\Bigg)\hat n_i \\&+ V^{sp}\sum_{i<j} \Bigg(\frac{\hat b^{\dagger}_{i}\hat b_{j}}{\abs{i-j}^3}+ \text{h.c.}\Bigg) \nonumber \\
	&= \frac{\Omega_{\mu w}}{2}\sum_{i}(\hat b^{\dagger}_i+ \hat b_i) - \sum_{i}(\Delta_{\mu w} + \mathcal{I}_i ) \hat n_i +\sum_{i<j}V_{ij} \hat n_i \hat n_j+\sum_{i<j} t_{ij}( \hat b^{\dagger}_i \hat b_j +\text{h.c.}),
\end{align}
where $V_{ij} = (V^s+V^p)/\abs{i-j}^6$, $t_{ij} = V^{sp}/\abs{i-j}^3$ and $\mathcal{I}_i=\sum_{i\neq j}\frac{V^{s}}{\abs{i-j}^6}$ with $V^s$, $V^p$ and $V^{sp}$ as defined in the main article. In the second line above, we dropped the constant term that comes from the fourth term in the first line. Dimerizing the above Hamiltonian gives
\begin{equation}
	\hat H _{dim}= \frac{\Omega_{\mu w}}{2}\sum_{i}(\hat b^{\dagger}_i+ \hat b_i) - \sum_{i}(\Delta_{\mu w} + \mathcal{I}_i ) \hat n_i +\sum_{i<j} \frac{C_3( \hat b^\dagger_i \hat b_j + \text{h.c.})}{(k_i a_1 + m_j a_2)^3} 
	+ \sum_{i<j} \frac{C_6(\hat n_i \hat n_{j} )}{(k_i a_1 + m_j a_2)^6},
\end{equation}    
where the distance between a pair of sites $(i,j)$ is given by $k_i a_1 + m_j a_2$ with $k_i,m_j \in \mathbb{N}$. Here, the lattice can be split into two sublattices consisting of odd and even sites. Writing nearest-neighbour terms for even and odd sublattices separately yields
\begin{align}
	\hat H_{dim} &= \frac{\Omega_{\mu w}}{2}\sum_{i}(\hat b^{\dagger}_i+ \hat b_i) - \sum_{i}(\Delta_{\mu w} + \mathcal{I}_i ) \hat n_i + \frac{C_3}{a_{1}^3}\sum_{i\in odd} ( \hat b^\dagger_i \hat b_{i+1} + \text{h.c.}) +\frac{C_6}{a_{1}^6}\sum_{i\in odd}\hat n_i \hat n_{i+1} \notag \\  &+ \frac{C_3}{a_{2}^3}\sum_{i\in even} ( \hat b^\dagger_i \hat b_{i+1} + \text{h.c.})+\frac{C_6}{a_{2}^6}\sum_{i\in even}\hat n_i \hat n_{i+1} + \sum_{\substack{i<j \\ k_i,m_j\neq 0}} \frac{C_3( \hat b^\dagger_i \hat b_j + \text{h.c.})}{(k_i a_1 + m_j a_2)^3} 
	+ \sum_{\substack{i<j \\ k_i,m_j\neq 0}} \frac{C_6(\hat n_i \hat n_{j} )}{(k_i a_1 + m_j a_2)^6}.
\end{align}
In the above equation, the sign of the $C_3$ coefficient can be changed by using a specific quantization axis. Here for the dimerized case, we set it to $-C_3/a_1^3$. Expressing all the interaction terms with respect to intra-cell interactions and defining $\alpha_l = a_1^3/(k_l a_1 + m_l a_2)^3$ with $\alpha \equiv \alpha_1(k_1=0,m_1=1)$ leads to the final form of the Hamiltonian
\begin{align}
	\label{eqn:H_dimfulll1}
	\hat H &=  \frac{\Omega_{\mu w}}{2}\sum_{i}(\hat b^{\dagger}_i+ \hat b_i) - \sum_{i}(\Delta_{\mu w} + \mathcal{I}_i ) \hat n_i +t\sum_{i\in odd} ( \hat b^\dagger_i \hat b_{i+1} + \text{h.c.}) + t\alpha\sum_{i\in even} (\hat b^\dagger_i \hat b_{i+1} + \text{h.c.}) 
	+ V\sum_{i\in odd} \hat n_i \hat n_{i+1} \notag \\ &+ V\alpha^2\sum_{i\in even} \hat n_i \hat n_{i+1}  + \underbrace{\sum_{\substack{i\\l=2}} t \alpha_{l} ( \hat b^\dagger_i \hat b_{i+l} + \text{h.c.})
		+ \sum_{\substack{i\\l=2}} V\alpha_l^2 \hat n_i \hat n_{i+l}}_{\hat H_{LR}},
\end{align}
where $t=-C_3/a_1^3$ and $V=C_6/a_1^6$. As mentioned in the main article, $C_6 = C^s_6 + C^p_6$ is the combined vdW dispersion coefficient. 
\begin{figure}[h!]
	\includegraphics[width=0.9\textwidth]{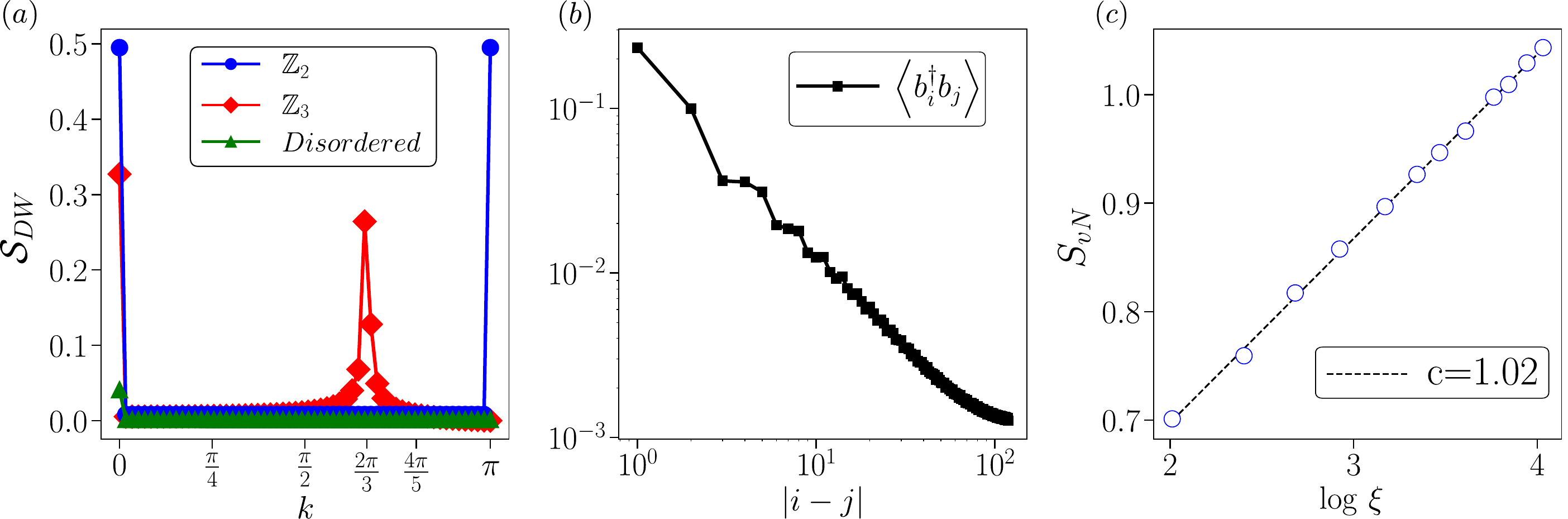}
	\caption{(a) Structure factor $\mathcal{S}_{DW}(k)$ for the phases shown in the main article  (Fig.~$2$(b)) showing pronounced peaks for the ordered phases but remaining featureless for the disordered phase (b) Hopping correlation function displays power-law decay behavior in the LL phase. (c) Scaling of $S_{vN}$ as a function of the correlation length $\xi$ in the LL phase. iDMRG simulations are performed with increasing bond dimensions to capture the scaling. The dashed line has been obtained via fitting according to the equation $ S = \frac{c}{6}\log(\xi) + $ const., where $ c $ is the central charge.}
	\label{fig:apx_uni}
\end{figure}

\section{Verification of phases for uniform lattice}
In this section, numerical verification of different phases observed in the uniform lattice is provided in the thermodynamic limit. Ordered (crystalline) phases are identified with the structure factor $\mathcal{S}_{DW}(k) = (1/L^2) \sum_{i,j}e^{ikr}\braket{\hat n_i \hat n_j}$. Crystalline phases exhibit long-range order which translates into sharp peaks of $\mathcal{S}_{DW}(k)$ at commensurate wave vectors $k = 2\pi n/q, n=0, \dots, q-1$ corresponding to DW order with $\mathbb{Z}_q$ translational symmetry breaking with $q=2,3$ as shown in Fig.~\ref{fig:apx_uni}(a). For the disordered phase,  $\mathcal{S}_{DW}(k)$ is featureless without any peaks. LL phase exhibits universal behavior such as power-law decay of correlations and the central charge $c=1$ . In Fig ~\ref{fig:apx_uni}(b) we show the hopping correlation function $\braket{\hat b^{\dagger}_i \hat b_j}$ which displays a power-law decay. We probed the growth of the entanglement entropy $S_{vN}$ as a function of the correlation length $\xi$ and extracted the universal central charge $c$ as displayed in Fig.~\ref{fig:apx_uni}(c).

\section{Verification of phases for dimerized lattice}
\begin{figure}[h!]
	\includegraphics[width=0.9\textwidth]{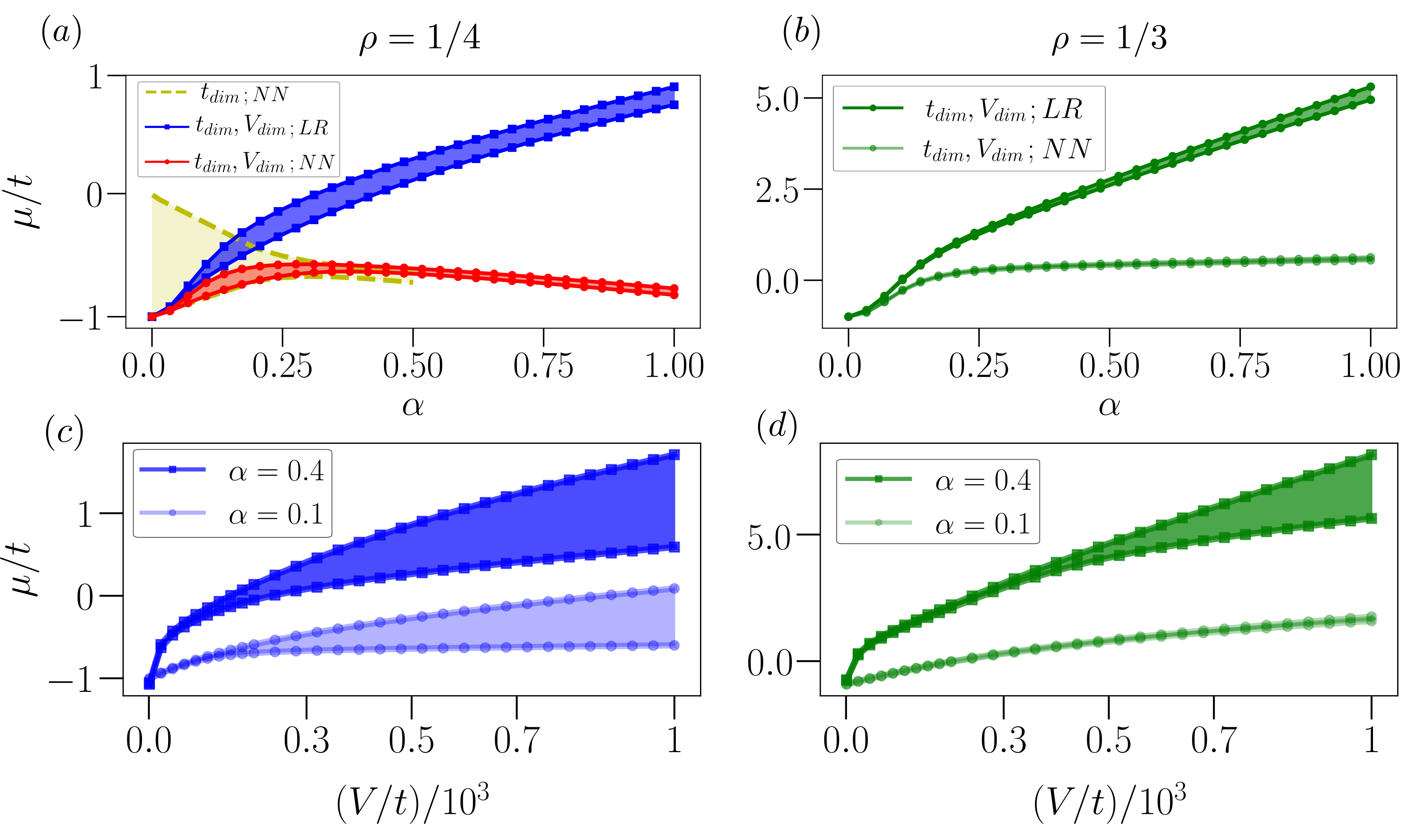}
	\caption{Comparison of BODW phases for different types of couplings (NN- Nearest Neighbour, LR- Long Range, $t_{dim}$- dimerized hopping, $V_{dim}$- dimerized interaction) in Eq.~$3$ in the main article with and without dimerization in the interaction for $\rho=1/4$ and $\rho=1/3$ as a function of $\alpha$ with fixed $V/t=200$ ($V/t=10$ for $t_{dim};NN$ case)  is shown in (a) and (b) respectively. (c,d) Comparison of BODW phases for different fixed dimerization values $\alpha=0.1, 0.4$ is shown as a function of $V/t$ for dimerized long-range interactions.}
	\label{fig:apx_dim}
\end{figure}

In this section, the numerical verification of different phases obtained in the dimerized lattice is provided in the thermodynamic limit and the BO to DW transition point is given. Analysis on BODW phases for different types of couplings and dimerization values is provided. To identify the phases, the structure factors and  expectation values of order operators are calculated. The BO is probed with the bond order structure factor $\mathcal{S}_{BO}(k) = (1/L^2) \sum_{i,j}e^{ikr}\braket{\hat B_i \hat B_j}$, where $\hat B_i = \hat b^\dagger_i \hat b_{i+1} + \text{h.c.}$ is the bond energy operator.
\begin{figure}[h!]
	\includegraphics[width=0.9\textwidth]{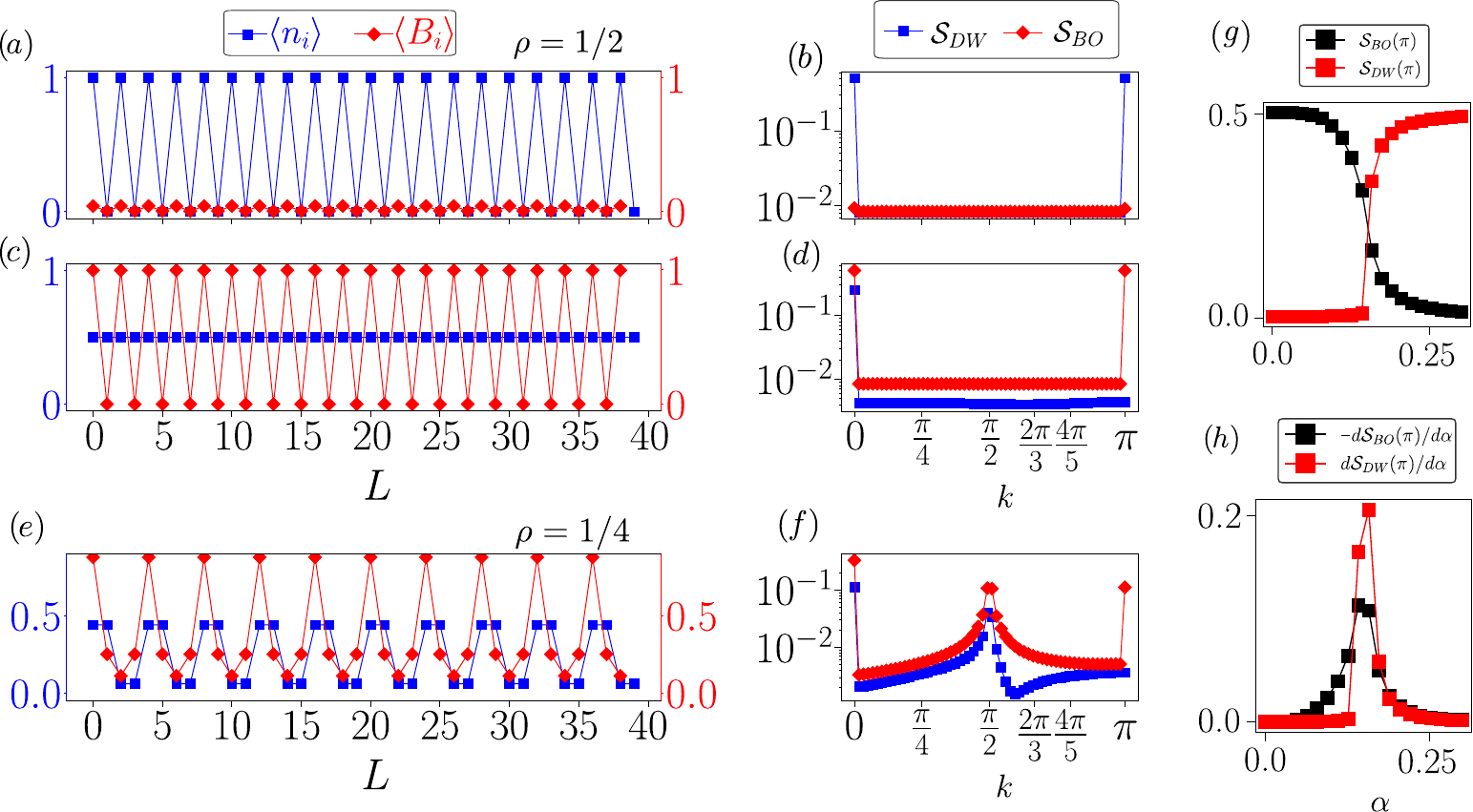}
	\caption{(a, c, e) Bond energy $\expval{B_i}$ and site density $\expval{n_i}$ expectation values for the gapped phases at each filling given in the main article (Fig.~$3$(a)) are displayed. Oscillations corresponding to DW, BO, and BODW orders are shown respectively. (b,d,f) Structure factors $\mathcal{S}_{DW}$ and $\mathcal{S}_{BO}$ exhibit pronounced peaks for the ordered phases. (g) $\mathcal{S}_{BO}(\pi)$ and $\mathcal{S}_{DW}(\pi)$ are determined as a function of $\alpha$. (h) Derivatives of $\mathcal{S}_{BO}(\pi)$ and $\mathcal{S}_{DW}(\pi)$ are given as a function of $\alpha$ .}
	\label{fig:dim_apx}
\end{figure}
In Fig.~\ref{fig:dim_apx} (a,c), oscillations in $\hat n_i$ and $\hat B_i$ corresponds to DW and BO phases at $\rho=1/2$ respectively. This oscillatory behavior is reflected in the peaks of  $\mathcal{S}_{DW}(k)$ and $\mathcal{S}_{BO}(k)$ at the wave vector $k=\pi$ as shown in Fig.~\ref{fig:dim_apx} (b,d). The critical point where BO to DW transition occurs is reached when $\alpha$ attains a value beyond which $\mathcal{S}_{DW}(\pi)$ is finite and $\mathcal{S}_{BO}(\pi)$ vanishes (Fig.~\ref{fig:dim_apx} (g)). In order to obtain the critical point, the derivative of $\mathcal{S}_{BO}(\pi)$ and $\mathcal{S}_{DW}(\pi)$ as a function of $\alpha$ is computed. The peak obtained for $d\mathcal{S}_{DW}(\pi)/d\alpha$ and $-d\mathcal{S}_{BO}(\pi)/d\alpha$ is around $\alpha \sim 0.16$ as shown in Fig.~\ref{fig:dim_apx} (h). Oscillations in both $\hat n_i$ and $\hat B_i$ (Fig.~\ref{fig:dim_apx} (e)) and sharp peaks of $\mathcal{S}_{BO}(k)$ and $\mathcal{S}_{DW}(k)$ (Fig.~\ref{fig:dim_apx} (f)) verify the coexistence of both DW and BO characteristics in BODW phases. For the BODW phase at $\rho=1/4$,  $\mathcal{S}_{BO}(k)$ makes a pronounced peak at $k=\pi/2, \pi$  and $\mathcal{S}_{DW}(k)$ at $k=\pi/2$ as shown in FIG \ref{fig:dim_apx} (f). This implies that the bosons are restricted to be found in every alternate unit cell, thus providing the DW character of the phase. Inside each alternating unit cell, bosons delocalize to minimize the ground state energy, which results in the bond formation. For the non-dimerized interacting model \cite{hayashi_competing_2022} shown with the yellow shade in Fig.~\ref{fig:apx_dim}(a), the gap vanishes as $\alpha$ increases because hopping is favoured causing delocalization of bosons. With dimerized nearest-neighbour interactions, we get a non-vanishing gap for small values of $\alpha$ but it gradually tapers off as shown by the red shaded region of Fig.~\ref{fig:apx_dim}(a). Finally, for the dimerized long-range interactions, one finds the gap to persist for finite values of $\alpha$ and is significant compared to the other two models. A similar analysis applies to Fig.~\ref{fig:apx_dim}(b) for a $\rho=1/3$ filling where it is necessary to have dimerization in the interactions in order to obtain the BODW phase. In Fig.~\ref{fig:apx_dim}(c,d), we analyze the effect of different fixed dimerization values on the BODW phases and showcase the importance of strong off-site interactions by varying $V/t$. The BODW phase at $\rho=1/4$ shrinks as the dimerization increases as shown in Fig.~\ref{fig:apx_dim}(c). Though, for sufficient $V/t$ the gap opens and gets larger as $V/t$ increases. In contrast with the previous case, the BODW phase at $\rho=1/3$ occurs only for $\alpha=0.4$ as shown in Fig.~\ref{fig:apx_dim}(d). This shows that beyond nearest-neighbor contributions play a more significant role when compared to the $\rho=1/4$ case.

\section{Numerical Methods}
In this section, details of the numerical calculations are presented. In this work, both finite and infinite matrix product states (MPS) are used for studying the ground state properties of the model in uniform and dimerized lattice configurations. All of the DMRG simulations are performed by using the TeNPy library \cite{hauschild_efficient_2018}. 

In the uniform lattice, a chain of length $L=121$ with open boundary conditions and units $a=1$,$V=1$ are adopted. The Hamiltonian is represented as a matrix product operator in which the power-law decaying dipolar and vdW interactions are expressed in terms of sum of exponentials. A decomposition that involves 15 exponentials is used to fit the interactions. The maximum MPS bond dimension is set to $\chi = 320$. We set the relative energy error to be smaller than $10^{-10}$ to ensure convergence. During truncation, Schmidt values smaller than $10^{-10}$ are discarded. Since open boundary conditions are employed, there will be defects induced by the edge excitations. In order to prevent that from happening, a system size in the form of $L=12n+1$ with $n=10$ is chosen. By doing so, the ground state $q$-fold degeneracy is split for phases with $\mathbb{Z}_q$ order with $q=2, 3, 4$. A single ground state in the product state form is obtained. This enables us to distinguish the ordered phases since they exhibit vanishing $S_{vN}$. The LL phase is verified by determining the central charge $c$ \cite{calabrese_entanglement_2009}. Infinite DMRG (iDMRG) simulations with increasing bond dimensions are performed to compute the scaling of the $S_{vN}$ vs. the logarithm of the correlation length $\xi$. The central charge $c$ \cite{pollmann_theory_2009} is extracted by a fitting according to the equation $ S = \frac{c}{6}\log(\xi) + $ const.

In the dimerized lattice, a system of $L=240$ sites with open boundary conditions and $t=1$  are used. The finite DMRG simulations are performed for obtaining the ordered regions whose boundaries indicate the closing of the single particle excitation gap $\delta_L$. The ground state for a system of chain length $L$ with $N$ bosons in the $\rho=1/2$ phase is obtained by considering a state $\ket{\Psi}_0=\ket{\circ \bullet \dots \circ \bullet}$ as an initial state. The ground state is then calculated by performing DMRG on this state while conserving the particle number $\sum_i^L n_i = N$. The energy for this ground state with $\braket{\sum_i^L n_i}=N$ is given as $E_L(N)$. The boundaries of the filling are determined by calculating the cost of creating a boson $\bullet$ or a hole $\circ$ in the system. For the upper (lower) boundary of the filling, the initial state for the DMRG is obtained by acting on $\ket{\Psi}_0$ with $\hat b^{\dagger}_i$($\hat b_i$) operators. Denoting the ground state energies for the particle and hole cases as  $E_L(N+1)$ and $E_L(N-1)$, the cost of boson creation can be given as $\mu_L^+ = E_L(N+1) - E_L(N)$ and for the hole creation $\mu_L^- =E_L(N) - E_L(N-1)$. This way the gap can be calculated and defined as $\delta_L=\mu_L^+ - \mu_L^-$. Similar calculations are performed in order to determine the boundaries of the gapped phases in other fillings. For the DMRG calculations, the maximum bond dimension is set to $\chi=350$ and a system of length $L=240$ is considered. After determining the lobes with ordered phases in both uniform and dimerized cases, iDMRG simulations inside these regions are performed to compute observables such as the structure factor and the expectation values of certain operators in the thermodynamic limit.

\end{document}